# High-performance fluoroelastomer-graphene nanocomposites for advanced sealing applications

Mufeng Liu[1], Pietro Cataldi[1], Robert J. Young[1], Dimitrios G. Papageorgiou[2*], Ian A. Kinloch[1*]

[1]*National Graphene Institute and Department of Materials, School of Natural Sciences, The University of Manchester, Oxford Road, Manchester M13 9PL, UK*

[2]*School of Engineering and Materials Science, Queen Mary University of London, Mile End Road, London E1 4NS, UK*

Corresponding authors: d.papageorgiou@qmul.ac.uk, ian.kinloch@manchester.ac.uk

## Abstract

High-performance sealing materials that can guard key components against high pressure gases and liquid chemicals while withstanding mechanical deformation are of utmost importance in a number of industries. In this present work, graphene nanoplatelets (GNPs) were introduced into a fluoroelastomer (FKM) matrix to improve its mechanical and barrier properties and test its suitability for sealing applications. Nanocomposites filled with different loadings of GNPs were prepared and compared with their counterparts loaded with carbon black (CB). GNPs were dispersed homogeneously with a high degree of in-plane alignment. The tensile and barrier properties of the FKM were improved significantly by the addition of GNPs. Micromechanical modelling based on the shear-lag/rule-of-mixtures theory was used to analyse the reinforcing efficiency of the GNPs. Upon the addition of the GNPs, the elastomer was able to swell anisotropically in liquids, a fact that can be used to tune the swelling properties for sealing applications. In terms of gas permeability, a modification of the well-established Nielsen's theory was introduced to analyse the $CO_2$ permeability for the bulk composite samples. The significantly improved mechanical, thermal and barrier properties at relatively low filler loadings, reveal that the FKM/GNP nanocomposites produced are very promising for use in advanced sealing applications.

# 1. Introduction

Due to the polar nature of their molecular composition, fluoroelastomers (FKMs) have good resistance to organic oils, enabling their use in a number of applications. They are currently used to reduce emissions during chemical processing and to facilitate power generation for the aerospace and automotive industries [1]. In addition, the curing systems of FKMs attribute good thermal stability and serviceability at elevated temperatures while the materials retain satisfactory flexibility and resilience at ambient temperature. Such characteristics are not as common in the elastomer family and therefore FKM components are used extensively in drive train and fuel handling systems (such as fuel injector O-rings, shaft seals, pump couplers, fuel-line hoses, etc.) [2]. For high-end applications under harsh environments, however, FKMs require improved properties in order to overcome the inevitable loss of their durability due to the oxidation of the network structure that leads to the reduction of mechanical and barrier performance [3, 4]. To meet the requirements of specific engineering applications, the introduction of inorganic fillers such as carbon black, silica and high-performing graphene-related nanomaterials (GRMs) have been considered as an effective approach, in order to create physical crosslinks and compensate for the network breakdown.

During the past decade, various elastomers have been used to produce multifunctional nanocomposites reinforced with graphene-related nanomaterials, where the mechanical [5-8], thermal [9-13], electrical [14-16], and barrier properties [17-22] have excited the scientific community as a result of their highly-promising performance. In terms of barrier properties, GRMs, possessing large aspect ratio, can efficiently create a tortuosity of pathways that prolong gas/liquid transportation in polymer matrices. A number of models [23-27] have been proposed to predict the barrier properties of filled polymer systems. Nielsen's approximation [25] quantifies the tortuosity of the gas diffusion path based on volume fraction and aspect ratio of platelet-like fillers that are assumed to be exfoliated perfectly. Bharadwaj [26] proposed a further development on the basis of Nielsen's model, where the orientation of the filler is also considered. Additionally, Cussler *et al.* [27] evaluated the gas permeability of thin nanocomposite films, where Nielsen's theory was modified for the case of low filler fractions. To date, most of the research on gas permeability of graphene-filled polymer systems has employed the aforementioned models directly, whereas the actual morphologies of the graphene-based fillers in bulk samples, include loops, folds, curves, stacks and agglomerates that are commonly formed during processing [5-7, 28, 29] and they have not been taken into consideration. Such complicated morphologies have been investigated using simulations and numerical calculations [24, 30-32], which, however, provide rather complex mathematical expressions that have not been adopted widely. Hence, it is necessary to propose a simple and effective gas barrier model that can quantify the contributions of the elements on the permeability phenomena in nanocomposite systems, and in which the actual geometries of the graphene-based fillers in bulk samples can be considered.

In this present study, we manufactured FKM nanocomposites filled with up to 15 phr of graphene nanoplatelets. An industry-standard elastomeric filler, carbon black (N990), was also utilised to produce nanocomposites with FKM, loaded at 15 and 30 phr for comparison of performance. The microstructure of the nanocomposites was examined using scanning electron microscopy (SEM). Tensile testing were conducted to characterise the mechanical properties of the materials. The reinforcing efficiency was investigated and both liquid and gas barrier properties were evaluated using swelling/mass diffusion and gas permeability experiments. The anisotropic swelling phenomenon was analysed and also explained comprehensively. Based on the actual microstructure of the nanocomposites, Nielsen's model [25] was adapted with the aid of Rutherford's mathematical route [24] to predict the gas permeability of the nanocomposites at relatively low filler fractions (semi-dilute regime).

## 2. Experimental methods

### 2.1 Materials and preparation

The fluoroelastomer (FKM), Tecnoflon® PFR 94,(Solvay Ltd.) was used as received. The graphene nanoplatelets (GNPs), received from Avanzare Ltd., coded as AVA-0240, with a surface area (BET) of 37 ± 12 (m$^2$/g) [33], were mixed with FKM. Thermax® N990 carbon black (CB) supplied by Cancarb Limited, Alberta, Canada was also mixed with FKM to create a second set of nanocomposites so that we could compare their performance with the FKM-GNP ones. The other additives employed in rubber processing for more effective mixing and curing, peroxide (Luperox® 101XL45, Arkema Co. Ltd.) and triallyl isocyanurate (TAIC 50 Co-activator, Wilfrid Smith, Ltd.) were of analytical grade and used as received. The detailed formulations of the samples produced in this work are listed in **Table 1**. Aditionally, acetone (anhydrous, 99.9%) (Sigma-Aldrich Co. Ltd.) was used to study the swelling of the elastomer nanocomposites.

*Table 1. Formulation of the FKM compounds*

| Materials | Loading (phr*) |
| --- | --- |
| FKM (Tecnoflon® PFR 94) | 100 |
| Peroxide (Luperox® 101XL45) | 1.5 |
| TAIC 50 Co-activator | 6 |
| GNP (AVA-0240) | 1, 2.5, 5, 10, 15 |
| N990 | 15, 30 |

*phr stands for part per hundred rubber by mass*

The FKM was compounded with the AVA-0240 at nominal loadings of 1, 2.5, 5, 10 and 15 phr and with N990 CB at nominal loadings of 15 and 30 phr in a two-roll mill at room temperature. The compounds prepared were then cut into pieces and compression moulded into sheets (~2.5 mm thick) using a Collin Platen Press (Platen Press P 300 P/M). The vulcanization took place at a temperature of 160 °C for 10 minutes under a hydraulic pressure of 30 bar. Afterwards, the hot-pressed sheets were post-cured in an oven at 230 °C for 2 hours. All the moulded elastomer sheets (~2.5 mm thick) were then stamped into dumbbell samples for tensile tests (ASTM D412) and disc-shaped samples with a diameter around 25 mm for the swelling and gas permeability tests. The processing procedure has been summarised into a flow chart that can be seen in **Figure S1**.

**2.2 Characterisation**

The actual loadings of the fillers in the nanocomposites were obtained by thermogravimetric analysis using a Jupiter® thermal analyzer (Netzch STA 449 F5). The transformations from the mass fraction to the volume fraction of the fillers in the nanocomposites were achieved using equation 1, with the densities of GNP (2.2 g/cm³), CB (2.0 g/cm³) and FKM (2.4 g/cm³) provided by the suppliers.

$$V_f = \frac{w_f \rho_m}{w_f \rho_m + (1-w_f)\rho_f} \tag{1}$$

where $w_f$ is the weight percentage obtained from TGA, $\rho_m$ and $\rho_f$ are the density of the matrix and the fillers, respectively. The samples were heated from room temperature up to 600 °C under a 50 mL/min flow of $N_2$ at 10°C/min. The sample mass, temperature and heat flow were continuously recorded in order to evaluate the thermal stability of the materials. At least 3 measurements were performed for every sample. Representative TGA curves of the samples are shown in **Figure S2**.

The microstructure of the nanocomposites was examined using scanning electron microscopy (SEM), after the samples had been cryo-fractured following immersion in liquid nitrogen. The cross-sectional surfaces of the samples were coated with 3 nm of Au/Pd alloy. The SEM micrographs were obtained using a TESCAN Mira 3 Field Emission Gun Scanning Electron Microscope (FEGSEM) operated at 5 kV. The lateral sizes of the GNPs in the nanocomposites were measured statistically through the SEM micrographs, using ImageJ software.

Tensile testing was conducted using an Instron 3365 with a load cell of 5 kN. At least 5 specimen were tested for each sample. The tensile tests were carried out using a cross-head speed of 500 mm·min⁻¹ in accordance with ASTM D412 (dumbbell).

For the liquid diffusion and swelling studies, the samples were immersed in ~50 mL of acetone. The uptake of the solvent was monitored by weighing the samples at intervals following immersion until their weight became constant and equilibrium was established. The dimensions of the samples were also measured using a vernier caliper in both the unswollen and equilibrium swollen states.

The carbon dioxide ($CO_2$) permeability was measured using a differential pressure testing system (**Figure S3**), in accordance with standard BS ISO 15105-1:2007. The gas permeability ($P$) was determined by the following equation:

$$P = \frac{V_{LP}}{R \times T_K \times p_{HP} \times A} \times \frac{dp_{LP}}{dt} \times h \tag{2}$$

where $V_{LP}$ is the volume of the low-pressure chamber, R is the gas constant ($8.31 \times 10^3$ J·K$^{-1}$·mol$^{-1}$), $T_K$ is the thermodynamic temperature, $p_{HP}$ is the pressure of the high-pressure chamber, $A$ is the transmission area of the specimen, $dp_{LP}/dt$ is the pressure change per unit time of the low-pressure chamber and $h$ is the thickness of the specimen.

## 3. Result and Discussion

### 3.1 Primary characterisation

The actual loadings of the prepared samples were initially examined using thermogravimetric analysis (TGA). The mass residue and volume fractions of the filler in the nanocomposites are given in **Table 2**. It can be seen that the experimentally-obtained mass fractions are close to the nominal mass fractions of the filler.

*Table 2. Mass residues and mass fractions of the fillers obtained by thermogravimetric analysis (TGA) and volume fractions of the fillers calculated using equation 1.*

| phr | Mass residue (%) | Mass fraction (%) | Volume fraction (%) |
|---|---|---|---|
| 0 | 8.30 ± 0.05 | 0 | 0 |

|     |     |                |                |                |
| --- | --- | -------------- | -------------- | -------------- |
|     | 1   | 9.71 ± 0.05    | 1.41 ± 0.10    | 1.54 ± 0.11    |
|     | 2.5 | 10.68 ± 0.10   | 2.38 ± 0.15    | 2.59 ± 0.17    |
| GNP | 5   | 12.51 ± 0.09   | 4.21 ± 0.14    | 4.57 ± 0.14    |
|     | 10  | 16.58 ± 0.15   | 8.28 ± 0.20    | 8.97 ± 0.22    |
|     | 15  | 19.93 ± 0.05   | 11.63 ± 0.10   | 12.55 ± 0.13   |
| CB  | 15  | 20.36 ± 0.05   | 12.06 ± 0.10   | 14.13 ± 0.15   |
|     | 30  | 29.44 ± 0.15   | 21.14 ± 0.20   | 24.34 ± 0.24   |

Scanning electron microscopy was employed to investigate the microstructure of the cross-sections of the prepared nanocomposites. As can be seen in **Figure 1** (a-f), the GNPs appear well-dispersed with their edges protruding from the cross-sections, while in **Figure 1** (g) and (h), the carbon black is present as approximately spherical nano-to-micron sized agglomerates. A high degree of in-plane orientation of the GNP flakes is generally found in the FKM matrix, as a result of the pressure applied on the rubber sheet plane during the compression moulding [7, 17, 18]. The GNPs can be seen well-wetted by the matrix, suggesting satisfactory filler/matrix interfaces. It is noticeable that curved, looped and folded geometries of the flakes are found to exist in the matrix, particularly for the samples with higher loadings (5, 10 and 15 phr) and for the relatively large flakes. This is in agreement with our previous findings for elastomer samples filled with GNPs provided by other suppliers [6, 7, 17, 18]. Such flake morphologies are observed as a consequence of the exfoliation method used to prepare the nanoplatelets that involves expansion of the graphite gallery and the processing methods that include high shear rates. A broad distribution of the lateral size of the flakes can be found in the SEM micrographs. Hereby, we have performed measurements of the flake sizes based on the SEM images, where large loops/folds have been excluded from measurements. The statistical results and representative loops/folds of the flakes can be found in Figure S4 - Supporting Information. Although from the SEM images it can be generally seen that there are a number of large flakes (~10 μm) presented in the matrix, the average lateral size of the flakes appears to be around 3 μm, based on the measurement of ~3000 flakes. This is due to the fact that there is a large number of small flakes (< 2 µm) that are present in the nanocomposite samples.

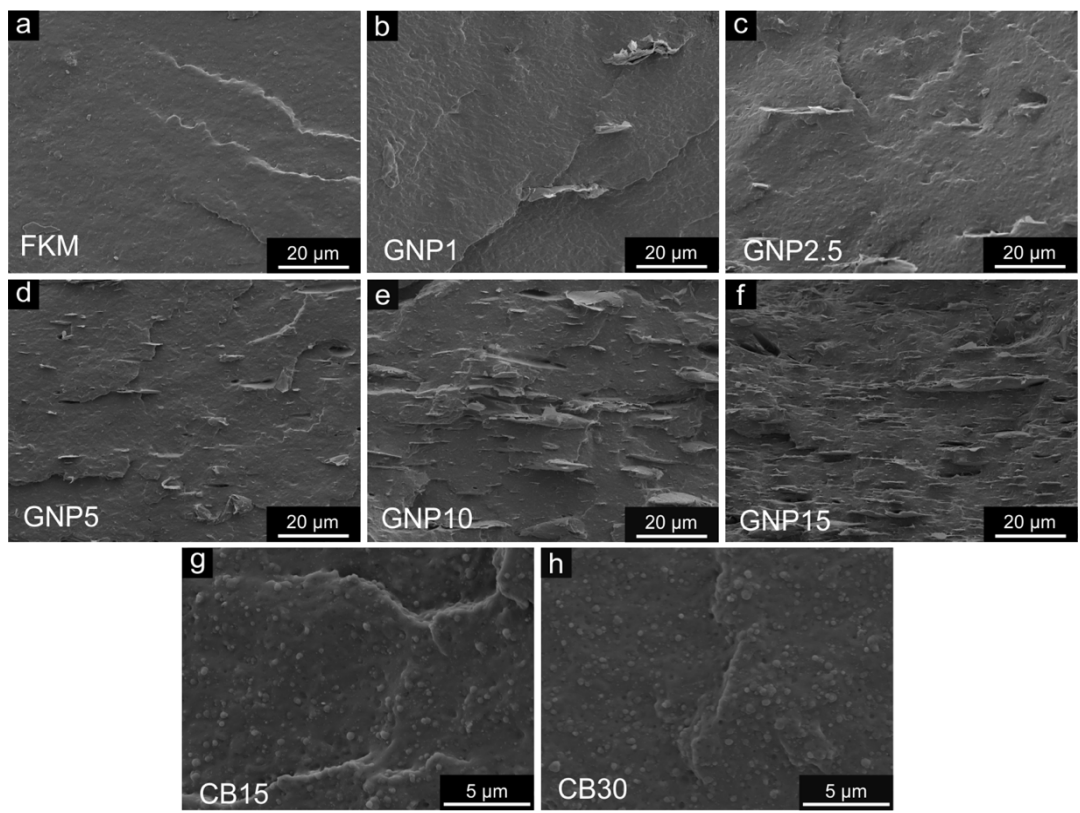

**Figure 1**. SEM micrographs for (a) neat FKM, (b-f) FKM filled with 1, 2.5, 5, 10 and 15 phr of GNPs, (g, h) FKM filled with 15 and 30 phr of carbon black.

**3.2 Mechanical properties**

The mechanical properties of the FKM nanocomposites were evaluated using tensile testing. The results are shown in **Figure 2**, where the volume fractions of the filler were confirmed by TGA (Figure S2), and presented in Table 2. It can be seen from **Figure 2** (a) that the tensile stress-strain behaviour of the FKM matrix was altered significantly by the addition of the GNP and CB. Both types of filler enhanced the stiffness and strength of the neat FKM. Compared with CB, the GNPs performed more efficiently in stiffening the elastomers due to their high aspect ratio that transfers the stress from the matrix to the filler and therefore reinforces the elastomer effectively [7, 8]. **Figure 2** (b) presents the normalized modulus at 50% strain, fitted by two micromechanical models, the shear-lag/rule-of-mixtures for GNP-filled samples that has been shown to explain accurately the reinforcement from 2D materials [8] and the Guth-Gold equation for CB-filled samples [34, 35], respectively. We opted to present the 50% modulus of the nanocomposites rather than the 100% in order to remain within the region of Gaussian deformation for all samples [36]. For the analysis of the modulus of the nanocomposites, we have shown previously that the reinforcing efficiency of GNPs in elastomeric matrices is dependent upon the aspect ratio and volume fraction of the filler, while it is independent of the filler modulus [5, 6, 8, 29, 37] . The normalized modulus of the composites is given by,

$$E_c/E_m = 1 - V_f + 0.056 s^2 V_f^2 \qquad (3)$$

where $E_c$ and $E_m$ are the moduli of the composite and the matrix, $s$ is the effective aspect ratio of the GNPs and $V_f$ is the filler fraction of GNP. Also for the CB-reinforced samples, the well-accepted Guth-Gold relationship employed here is

$$E_c/E_m = 1 + 0.67 f V_f + 1.62 f^2 V_f^2 \qquad (4)$$

where $f$ is the aspect ratio (length/breadth) of a rod-like shape that can be formed by agglomerated CB nanoparticles. Here, the $f$ was determined to be 3.5 for the CB samples ($f \leq 6$ has been empirically accepted for CB filled elastomers) [33]. It can be seen both equations fit effectively the experimentally obtained modulus values, where for GNPs, the effective aspect ratio ($s_{eff}$) was evaluated to be in the order of 80. The tensile strengths of the samples are shown in **Figure 2** (c). The tensile strength of the GNP-filled samples reached a plateau at relatively low filler loadings (2.5 phr) since high filler loadings of GNPs can lead to the inevitable formation of aggregates. The GNPs and CBs both show almost the same tensile strength at 15 phr loading. Unlike the Young's modulus of the nanocomposites, the statistical nature of tensile strength is much more sensitive to the formation of large aggregates that can act as failure points during the elongation of the samples and can reduce the ultimate tensile strength. Additionally, as a result of the Payne effect [38-40], the applied tensile strain breaks down the weakly bound carbon black agglomerates at high CB contents, leading to an increase of the tensile strength due to the better dispersion of the filler. The failure strains of the samples are shown in **Figure 2** (d). It can be seen that the addition of the GNPs up to 10 phr did not change the failure strains of the materials significantly. With the addition of CB at 15 phr and 30 phr, the elongation showed similar values (~300%) with the sample filled with 15 phr GNP. It can be concluded, therefore, that the FKM samples with the addition of either GNP or CB still exhibit remarkable elastomeric characteristics, which are crucial for the commercial application of FKMs.

Overall, for the key mechanical properties characterised here, it is worth pointing out that the samples reinforced by 2.5 phr GNPs display properties similar to the ones of the samples filled with 15 phr CB, indicating that the GNPs can be much more efficient than CB at improving the mechanical properties of the rubber at lower filler contents.

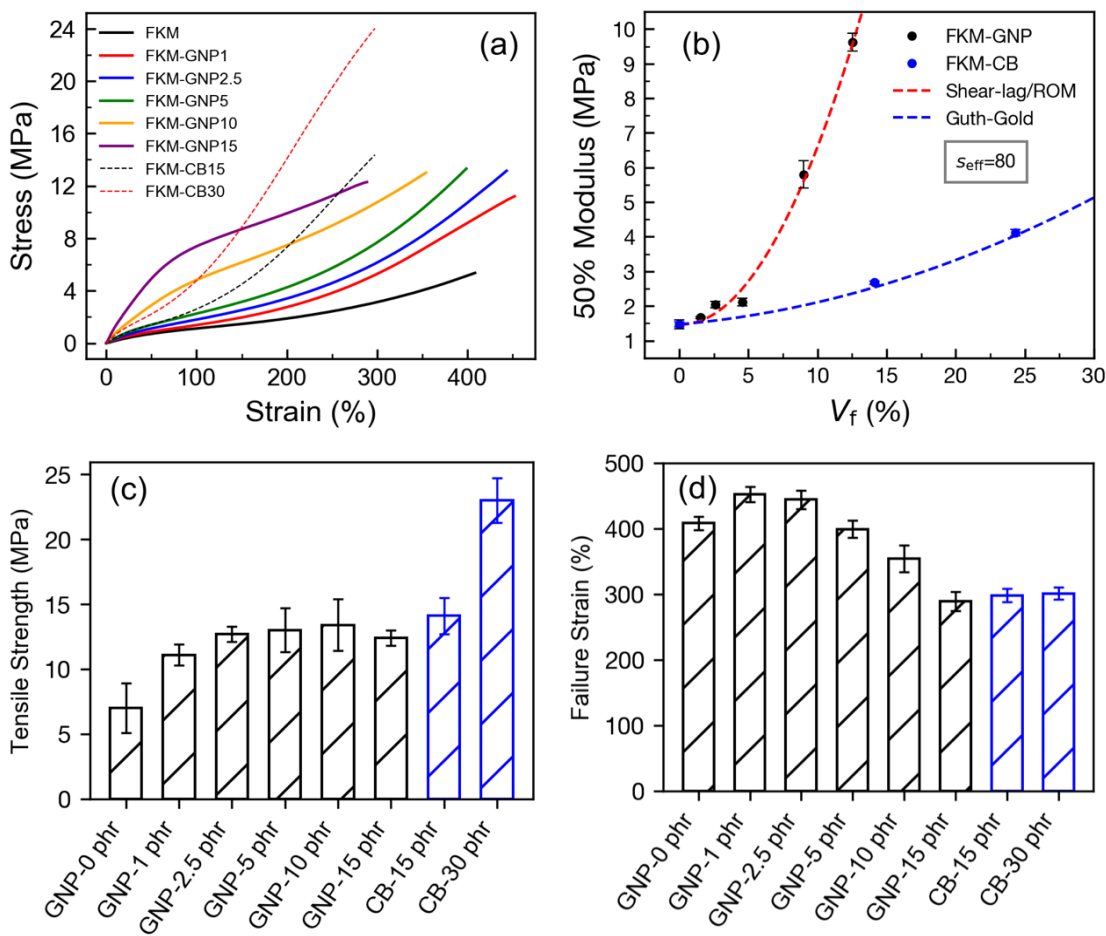

**Figure 2**. Mechanical properties of the FKM nanocomposites filled with GNPs and CB. (a) Representative stress-strain curves of the materials under tensile testing; (b) Modulus of the samples

at 50% strain for different loadings of GNPs and CB, fitted by shear-lag/ROM theory (GNP) and Guth-Gold theory (CB); (c) Failure strains of the samples.

**3.3 Diffusion and swelling in acetone**

FKM possesses excellent liquid barrier properties to various solvents, particularly to non-polar oils, as a result of its polarity and molecular structure. Herein, we aimed at improving the liquid barrier properties of FKM even further by using relatively low filler fractions, therefore widening its potential sealing applications. The FKM and the nanocomposite specimen were immersed in acetone and continuous measurements of the weight of the swollen specimen at regular time intervals were conducted in order to investigate the relative mass uptake $M$, and the diffusion coefficient $D$. The relative mass uptake $M(t)$ was determined gravimetrically as a function of the exposure time by $M(t)=\frac{W(t)-W(0)}{W(0)}$, [41] where $W(t)$ is the weight of a specimen after an exposure time $t$ and $W(0)$ is calculated theoretically by determining the intercept of the linear fitting of $W(t)$ against $t^{1/2}$ at $t=0$, in order to reduce any systematic error involved in gravimetry.

It has been shown [41] that the diffusion coefficient, $D$, for a polymer specimen with a length $a$, width $b$ and height $h$ for short exposure times $t$, can be obtained by,

$$\frac{M(t)}{M(\infty)}=1-\left(1-\frac{4}{h}\sqrt{\frac{Dt}{\pi}}\right)\left(1-\frac{4}{a}\sqrt{\frac{Dt}{\pi}}\right)\left(1-\frac{4}{b}\sqrt{\frac{Dt}{\pi}}\right) \qquad (5)$$

and for a plate-like specimen (with a diameter of at least 10× the thickness), that is, with the in-plane dimensions $a\rightarrow\infty$ and $b\rightarrow\infty$, equation (5) can be reduced to the following form for the experimental determination of the diffusion coefficient, $D$,

$$\frac{4M(\infty)}{h\sqrt{\pi}}\sqrt{D}=\frac{M_2-M_1}{\sqrt{t_2}-\sqrt{t_1}} \qquad (6)$$

which allows us to evaluate the diffusion coefficient from the mass uptake and the thickness of the disc-shape samples.

The relative mass uptake $M(t)$ as a function of $t^{1/2}$ for the FKM/GNP and the FKM/CB nanocomposites are shown in **Figure 3** (a) and (b). At the beginning of the exposure to the liquid environment, the small molecules of the solvent diffused into the materials leading to an increase of $M(t)$. Afterwards, the plots started to plateau at $t^{1/2} \approx 100$ ($s^{1/2}$), which indicates the saturation of absorption. When neat FKM was exposed in the solvent, the small molecules of the solvent diffused into the free volume of the rubber, making it swell in three dimensions, until a balance (equilibrium) between the deformation (swelling) of the polymer network and the osmotic pressure of the solvent was achieved. When the fillers were incorporated into the matrix, however, the internal stress resulting from the swelling of the matrix can be transferred from the matrix to the filler through shear at the GNP/matrix interface, which eventually restrained the swelling of the elastomer network and therefore reduced the overall mass uptake of the materials. Therefore from both **Figure 3** (a) and (b), it can be seen that $M(\infty)$ was reduced with increasing filler content for both GNP and CB filled samples. The diffusion coefficients for all the samples were calculated using equation (6) and shown in **Table 3**. It is demonstrated that the GNPs, owing to their high aspect ratio again, were able to increase the tortuous path of the liquid much more efficiently that CB and therefore their swelling rate was much slower.

*Table 3. The diffusion coefficients for all the samples immersed in acetone.*

| FKM-GNP | | FKM-CB | |
|---|---|---|---|
| phr | $D$ ($\times 10^{-5}$) (mm$^2$/s) | phr | $D$ ($\times 10^{-5}$) (mm$^2$/s) |
| 0 | 14.2 ± 0.01 | 0 | 14.2 ± 0.01 |
| 1 | 14.1 ± 0.02 | 15 | 11.5 ± 0.01 |
| 2.5 | 12.1 ± 0.01 | 30 | 11.3 ± 0.01 |
| 5 | 9.8 ± 0.01 | | |
| 10 | 7.8 ± 0.03 | | |
| 15 | 7.6 ± 0.01 | | |

Another important and interesting observation from this experiment is the anisotropic swelling for the GNP-reinforced samples. The volumetric and dimensional swelling were both recorded when the samples were saturated in the solvent (at the equilibrium). It can be seen in **Figure 3**(c) that the diameter and volume swelling both decreased with increasing GNP loadings, while counterintuitively, the thickness swelling increased. This phenomenon was revealed and modelled in detail in our previous studies [17, 18]. The qualitative explanation of this phenomenon can be traced to the swelling of rubber under a tensile force. Treloar [36, 42] has described the increase of swelling of a rubber under a tensile force, where the tension applied to the rubber distorts the network structure and increases the free volume of the network along the direction of the applied force. In this present work, the anisotropic swelling was essentially caused by the high degree of orientation of the GNPs (shown in **Figure 1**) due to the compression moulding. Since the compression moulding procedure resulted in an in-plane orientation of the GNPs, it can easily be understood that the polymer network close to the interfaces is also oriented in-plane. In the in-plane direction, the swelling can be hindered by the GNPs, compared with the unfilled rubbers. In the out-of-plane direction, however, the rubber network can display much larger free volume, given that the internal structure of an unfilled elastomer can be approximated to a 3D random network. The distorted network structure subsequently led to solvent segregation during the swelling, leading to reduced diameter swelling and increased thickness swelling, with increasing filler fractions.

The results of the anisotropic swelling can be modelled quantitatively based on the Flory-Rehner-Treloar statistical mechanics theories [18, 36, 42-44]. The diameter and thickness swelling ratio are given by [18],

$$d_e/d_0 = (V_e/V_0)_{neat}^{-5/12} \cdot (V_e/V_0)^{3/4} \tag{7}$$

$$h_e/h_0 = (V_e/V_0)_{neat}^{5/6} \cdot (V_e/V_0)^{-1/2} \tag{8}$$

where $(V_e/V_0)_{neat}$ and $(V_e/V_0)$ are the volume swelling ratios of the neat FKM and the nanocomposites, respectively, while $d_e/d_0$ and $h_e/h_0$ are the diameter and thickness swelling ratio, respectively. For CB-filled samples, the swelling of the samples is isotropic and therefore [36, 43],

$$d_e/d_0 = h_e/h_0 = (V_e/V_0)^{1/3} \qquad (9)$$

The experimental results of the diameter and thickness swelling were fitted using equations (7) and (8) and shown in **Figure 3**(d) where the *x* axis represents the values of the $V_e/V_0$ of the samples, which can be seen in very right (blue) *y* axis of **Figure 3**(c). The filler loadings of the samples increases from higher to lower values of $(V_e/V_0)$, as indicated by the arrows. The products $d_e/d_0$ (left/black Y axis) and $h_e/h_0$ (right/red Y axis) refer to the swelling ratio of the in-plane (diameter) and out-of-plane (thickness) directions, respectively. As suggested by **Figure 3**(c), with the increase of the filler loading, both $d_e/d_0$ and $V_e/V_0$ decreased, and $h_e/h_0$ increased. It can be seen that the experimental data points lie on the lines predicted by the theory (eqs 7 & 8) consistently. An important conclusion that can be exported from the fittings is that the overall filler orientation within the nanocomposites can be considered to be in-plane, given that the theoretical curves (eqs. 7 & 8) plotted in **Figure 3**(d) are based on the assumption of perfect in-plane orientation of GNPs [18]. This useful conclusion will be also utilized for the analysis of the gas permeability in the next section, where we can ignore the effect of orientation of the 2D materials for the gas permeation of the samples studied in this work.

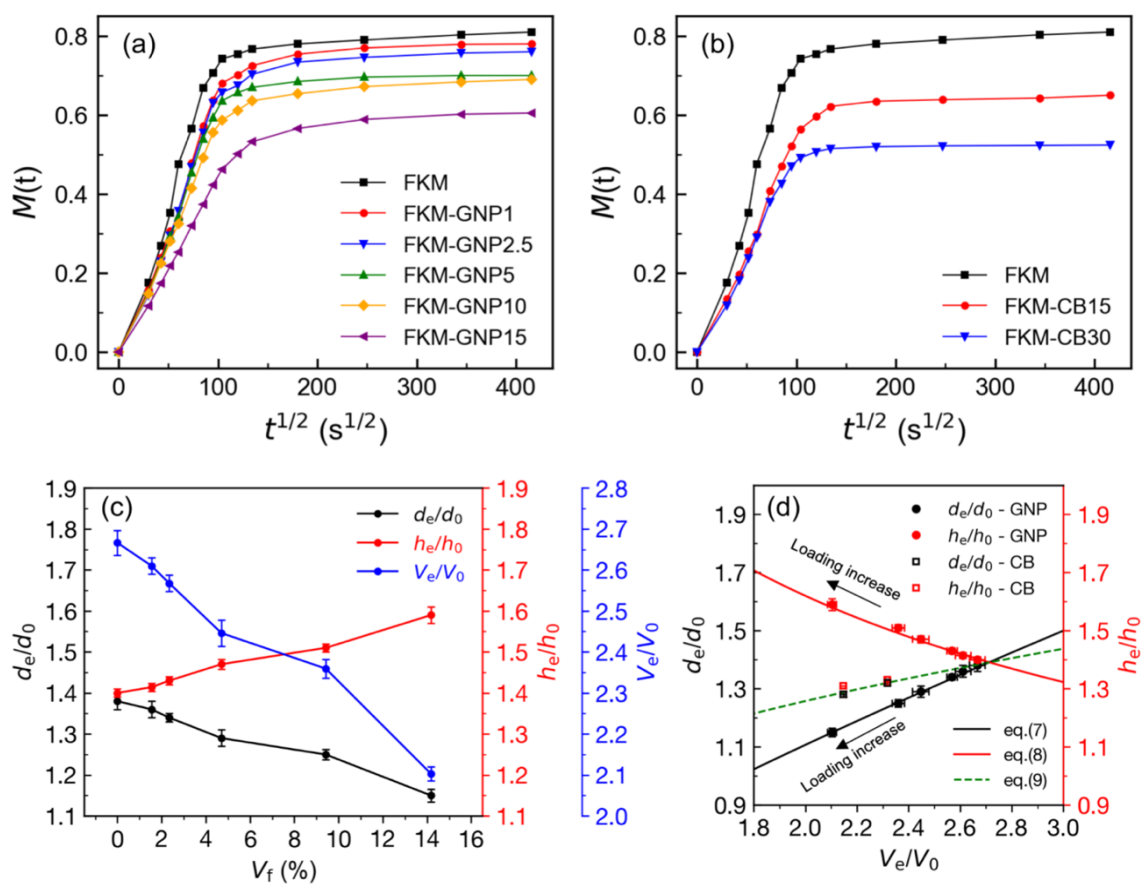

**Figure 3**. Mass uptake of acetone for FKM and its nanocomposites filled with (a) GNP and (b) CB; (c) dimensional swelling at the equilibrium for GNP-filled samples against the volume fraction of GNPs and (d) theoretical analysis of the diameter and thickness swelling using equations 7, 8 and 9.

### 3.4 Gas barrier properties

Given the unique high-temperature serviceability and low-temperature flexibility of the FKMs, they can be used extensively as seals for gas tanks used in hybrid automobiles (such as bi-fuel systems with petrol and natural gas) that emit less $CO_2$ than conventional cars. Hence, the gas barrier properties of FKM and its nanocomposites is another crucial factor to be evaluated for use in advanced sealing applications. We therefore characterised the $CO_2$ permeability of the nanocomposites to examine the advantages of (in-plane-aligned) GNPs over CB in gas permeation hindrance. The measured gas pressure (in the lower pressure chamber $p_{LP}$ – the schematic of the gas measurement setup can be seen in Figure S3 of Supporting Information) against time is shown in **Figure 4**(a) and the calculated permeabilities of the materials are shown in **Figure 4**(b) and tabulated in Table S2. As can be observed from **Figure 4**(a), on the basis of equation (2), the reduced slope of the pressure in the $p_{LP}$ chamber with time indicates the reduced permeability of the samples with increasing GNP loadings. It is demonstrated that the permeation of $CO_2$ through FKM was reduced significantly by the addition of GNPs (50% reduction by 15 phr GNPs) that performed clearly better than CB. The mechanism of the reduction of the gas permeability by the GNPs is considered to be the increased tortuosity of the paths for the gas molecules. During the permeability measurements, the gas was transported in the out-of-plane direction through the elastomer sheets, which is perpendicular to the orientation of the GNPs (in-plane). The high degree of orientation of the GNPs (confirmed in sections 3.1 and 3.3) can maximize the performance of the nanoplatelets on the permeability reduction of the nanocomposites [26]. The carbon blacks, on the other hand, cannot impose an equally efficient barrier as a result of their spherical shape and low aspect ratio.

We have evaluated theoretically the efficiency of the GNPs towards the gas barrier properties of the GNP-reinforced samples by modifying the well-established Nielsen's model [25]. Nielsen's theory is arguably the most widely-used theory to describe the permeability of polymer nanocomposites. The derivation of Nielsen's equation is based on the assumption that the in-plane end-to-end distance between flakes is infinitely low. However, for low loadings such as the ones used in this present study, as it can be also seen from the SEM micrographs (**Figure 1**), this is not true. Therefore, the transport path is unlikely to be maximised, as Nielsen's model suggests. Hereby, we employ a mathematical expression similar to Nielsen's one, in order to derive a new equation, where the various flake sizes and the relatively low filler loading are considered. The detailed derivation of the equations are presented in **Section S6-Supporting Information**.

In the Nielsen model [25], the permeability of the composite ($P_c$) is given by,

$$P_c = P_m \left( \frac{1-V_f}{\tau} \right) \tag{10}$$

where $P_m$ is the permeability of neat polymer and $V_f$ is the volume fraction of the filler. In this model, the tortuosity factor ($\tau$) is defined as the ratio of the actual distance ($d_a$) that a penetrate travels, to the shortest distance ($d_s$) that it would have travelled in the absence of the platelet filler, and is given by,

$$\tau = \frac{d_a}{d_s} = 1 + \frac{s}{2} V_f \tag{11}$$

where $s$ is the aspect ratio of the filler. Nielsen's equation is an approximation of the lower bound of the permeability of the nanocomposite materials, where the gas permeation takes place in the direction perpendicular to aligned, homogeneously-dispersed and perfectly-exfoliated platelets, and therefore the tortuosity of the gas transport is maximized, as can be seen in the schematic diagram of **Figure 4**(c). For the present work, the GNPs appeared to display various lateral sizes with some unexpected geometries (loops and folds) within the composites (as discussed in Section 3.1) as a result of the manufacturing method and subsequent processing under high shear. In this case, the minimum permeability theory proposed by Nielsen is not applicable to the analysis of bulk GNP/elastomer nanocomposites (with relatively low loadings, up to ~15 vol%).

On that basis, we have adapted the tortuosity factor to account for the gas barrier performance of polymer nanocomposites reinforced by 2D materials. As can be seen in the schematic diagram in **Figure 4**(d), the GNPs mixed into the FKM matrix display a size distribution. Therefore, the gas molecules do not encounter a large number of flakes through the thickness of the sample. As a result, the increase of tortuosity due to the presence of the 2D materials is limited to a certain extent. Hereby, we can divide the composite materials into many layers ($N$ layers in total), where each individual layer contains a thickness of a flake ($t$) and an amount of interphase with a thickness of $T$. In the vertical direction (out-of-plane) of the model the total thickness of the material is $N(t+T)$. For a gas molecule travelling from the surface to the bottom through the thickness of the composite, it can either encounter with or miss a flake. It is assumed that in the case when the gas molecule encounters with the flake, the flake can increase its travel distance by $l/2$, similar to the Nielsen model. If we assume there are $n$ layers where a gas molecule encounters with a flake throughout its travel, then there are ($N$-$n$) layers where this gas molecule has missed a collision with the flakes. The actual distance of the travel of a penetrate is therefore given by,

$$d_a = d_m + d_h \tag{12}$$

where $d_m$ is the travel distance when the penetrate skipped flakes, which is ($N$-$n$)·($t$+$T$), and $d_h$ is the travel distance when the penetrate encounters flakes, which is $n$·($t$+$T$+$l$/2). The shortest distance that the molecule would have travelled in the absence of the filler, $d_s$ is,

$$d_s = N \cdot (t+T) \tag{13}$$

Hence, the tortuosity factor for this model is,

$$\tau = \frac{d_a}{d_s} = \frac{N(t+T)+nl/2}{N(t+T)} = 1 + \frac{n}{N} \cdot \frac{l}{2(t+T)} \tag{14}$$

Subsequently, it is important to evaluate the possibility of a gas molecule colliding with $n$ flakes. Since the value of $N$ is large, the parameter $n$ can take its expected (mean) value, $\bar{n}$, for the calculation of the tortuosity factor. The possibility ($P_n$) that the gas molecule can encounter $n$ flakes can be obtained by a binomial distribution [24, 27], which subsequently gives,

$$V_f = \frac{\bar{n}}{N} \tag{15}$$

where $V_f$ is the volume fraction of the filler. The detailed mathematical derivation of equation (15) can be found in Section S6 – Supporting Information. A similar mathematical approach was developed by Rutherford [24] and employed by Cussler *et al.* [27]. Cussler's equation is given by:

$$P_c = P_m \cdot \frac{1-V_f}{1+\frac{s^2}{4}V_f^2} \text{ [27]}.$$

The product, $\frac{l}{2(t+T)}$, in equation (14) can be multiplied by the flake thickness ($t$) in both the numerator and the denominator, which then gives $\frac{1}{2} \cdot \frac{l}{t} \cdot \frac{t}{(t+T)}$. Equation (14) can therefore be simplified into

$$\tau = 1 + \frac{s}{2} \cdot V_f^2 \tag{16}$$

where the aspect ratio ($s$) is equal to the length/thickness ($l/t$) of the flakes and the volume fraction of the filler ($V_f$) is approximated to $t/(t+T)$.

Hence, the permeability of the bulk composite ($P_c$) can be expressed to be,

$$P_c = P_m \cdot \frac{1-V_f}{1+\frac{s}{2} \cdot V_f^2} \tag{17}$$

where $P_m$ is the permeability of the neat matrix. The tortuosity factor derived here, $(1+\frac{s}{2} \cdot V_f^2)$, is very similar to Nielsen's minimum permeability model, $(1+\frac{s}{2} \cdot V_f)$. The only difference is that our model takes into account the possibility of hitting or missing the flakes when a gas molecule travels through, in order to adapt to the case of a bulk composite reinforced by low filler loadings, which subsequently added a power of 2 onto the volume fraction. As can be seen in **Figure 4(e)**, the adapted model fits the experimental results excellently by considering an effective aspect ratio of 80, consistent with the analysis of mechanical properties.

Theoretically, the upper bound of gas permeability within a composite sample can be considered by only taking into account the contribution of the volume fraction of the impenetrable filler, where there is no tortuosity generated ($\tau = 1$) and therefore $P_c/P_m = (1-V_f)$. The experimental results from carbon black filled samples were consistent with this case, as can be seen in **Figure 4(e)**. Comparing the adapted model proposed herein with the Nielsen model and Cussler model from 0 to 100% filler fraction in **Figure 4(f)**, it can be realised that when the filler loadings reach 60 vol%, all the models predict similar permeabilities, since larger filler volume fractions cannot improve the permeability of the materials significantly. Additionally, it can be clearly observed that on the basis of our model that takes into account the contribution from flakes with a variable size distribution and flake characteristics that lead to the reduction of the effective aspect ratio (such as loops and folds), the permeability decreases at much slower rates, compared to Nielsen's and Cussler's models. It is worthwhile to mention that the model proposed in this work only accounts for the tortuosity of the path created by oriented 2D materials within elastomers (which display very low values of crystallinity, if any). For semicrystalline polymers filled with nanomaterials, the change of their crystallinity as a result of the presence of nanofillers is known to be a crucial factor that affects permeability [45, 46]. Functionalized graphene or graphene oxide that can induce a further improvement of the filler/matrix interfacial adhesion also enables further reduction of the permeability of the nanocomposites [47, 48] that cannot be predicted by this model.

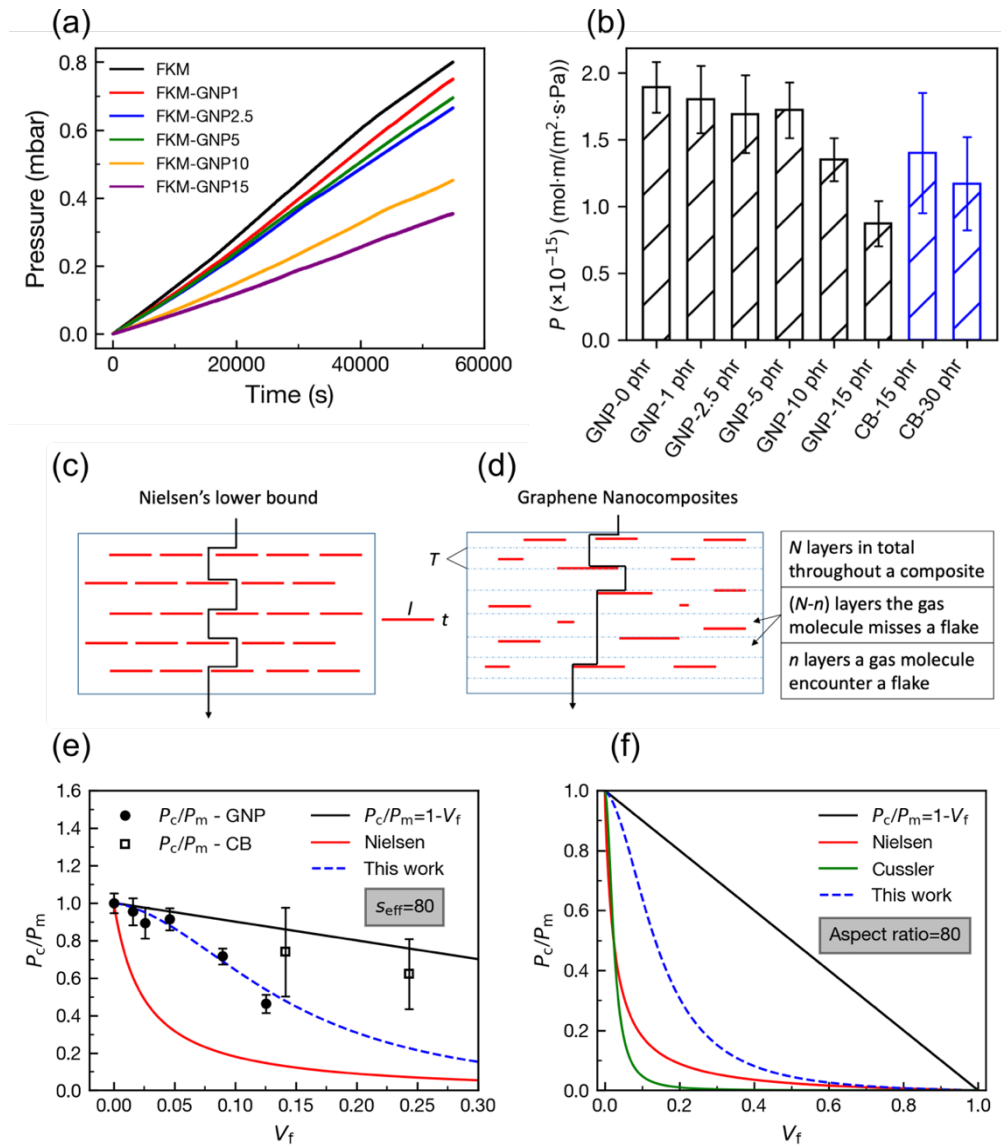

**Figure 4.** Measured results of (a) pressure of the lower pressure chamber ($p_{LP}$) against time and (b) permeabilities for all the samples; (c) Nielsen's of the path of gas transportation predicting the minimum permeability; (d) Schematic diagram of the path of gas transportation adapted for this work where a size distribution of the flakes at low loadings of the filler is considered; the model enables the analysis for the gas molecules to miss a flake. (e) curve fitting with eq.18 (dashed blue curve) on the permeability results for GNP-filled samples, while the effective aspect ratio of 80 was obtained.

The upper bound (1-$V_f$) (black curve) and lower bound (Nielsen model, red curve) are also shown with $s_{eff}$ equal to 80 set for the Nielsen model); (f) comparison of models (GNP aspect ratio=80) for volume fraction of the filler from 0 to 1 (0%-100%), including Nielsen's model, Cussler's model and the model developed in this work.

## 4. Conclusions

Fluoroelastomer/GNP nanocomposites were successfully prepared using a two-roll-mill method. The SEM studies revealed that the prepared FKM/GNP nanocomposites contained a large number of in-plane oriented GNPs, due to the high pressure applied to the rubber sheets during the compression moulding process. The GNPs, owing to their large aspect ratio along with the in-plane orientation, contributed to better mechanical properties, anisotropic swelling behaviour, while they also restrained liquid diffusion and reduced the gas permeability of the nanocomposites to a larger extent than their carbon black counterparts.

The reinforcement of the FKM/GNP nanocomposites was evaluated using micromechanics, revealing an effective aspect ratio of 80 for the GNPs. A novel equation on the basis of the Nielsen's permeation model, was derived and used to analyse the permeability of the FKM/GNP samples, taking the lateral sizes and low content of the GNPs into account. It was revealed that an effective aspect ratio of 80 was again able to fit the data satisfactorily, in agreement with the analysis of the mechanical properties using micromechanics. The measurement of the lateral dimensions of the flakes from SEM images suggested an estimated lateral size of ~3 μm (Figure S4). Based on the specific surface area of 37 $m^2$/g reported by the manufacturer of the nanoplatelets [33], the theoretical aspect ratio of the flakes can be estimated to be in the order of 200-300. If the large loops, folds and agglomerates which reduce the effective aspect ratio of the flakes are taken into account, it is plausible that the effective aspect ratio of the flakes contributing to the properties is ~80, as concluded from both theories.

Overall, it can be concluded that the improvements of the mechanical and barrier properties from the GNPs rely on the effective aspect ratio of the flakes. The processing method affects the dispersion and orientation significantly, which determines the final filler geometry in the bulk nanocomposites. The properties of the GNP nanocomposites were clearly better than the CB nanocomposites, providing anisotropic properties as well, which can offer the possibility to tune the properties of the prepared nanocomposite materials for targeted applications. For a number of sealing applications where FKM elastomers are currently used, particularly for the automotive industry, GNPs show great potential to be employed as highly effective reinforcements, contributing greatly towards improving the performance and durability of the advanced seals.

**Acknowledgements**

The authors acknowledge the support from "Graphene Core 2", GA: 785219 and "Graphene Core 3" GA: 881603 which are implemented under the EU-Horizon 2020 Research & Innovation Actions (RIA) and are financially supported by EC-financed parts of the Graphene Flagship. Ian A. Kinloch also acknowledges the Royal Academy of Engineering and Morgan Advanced Materials. All research data supporting this publication are available within this publication.

# High-performance fluoroelastomer-graphene nanocomposites for advanced sealing applications

Mufeng Liu[1], Pietro Cataldi[1], Robert J. Young[1], Dimitrios G. Papageorgiou[2*], Ian A. Kinloch[1*]

[1]National Graphene Institute and Department of Materials, School of Natural Sciences, The University of Manchester, Oxford Road, Manchester M13 9PL, UK

[2]School of Engineering and Materials Science, Queen Mary University of London, Mile End Road, London E1 4NS, UK

Corresponding authors: d.papageorgiou@qmul.ac.uk, ian.kinloch@manchester.ac.uk

*Supporting Information*

## S1. Experimental procedures

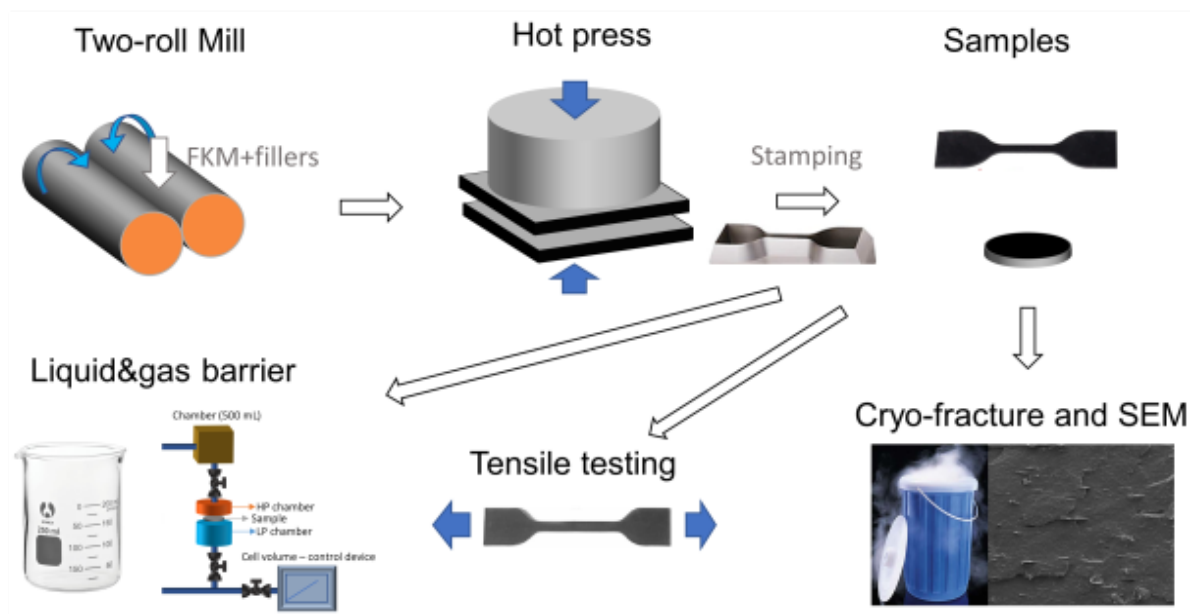

**Figure S1**. The experimental procedures of the processing and characterisations.

## S2. Thermogravimetric analysis

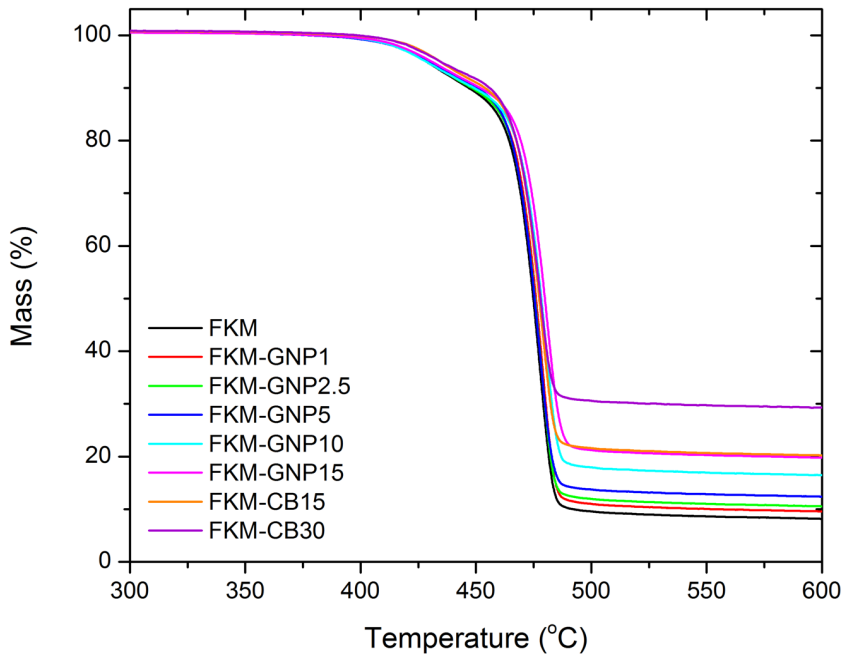

**Figure S2**. Thermogravimetric analysis (TGA) curves of all the samples.

## S3. Gas permeability rig

The testing system (**Figure S3**) was initially stabilised under vacuum for 24 hours prior to the measurement. When the testing unit (between valve 4 and 5) was under full vacuum, valve 1 was opened, while all the other valves (2-6) were closed, in order to allow the gas to fill the chamber (with a fixed volume of 500 mL) from the high pressure (HP) gas cylinder. Once the 500 mL chamber was fully filled, valve 1 was closed, followed by the start of the measurement (valve 4 opened). The pressure in the low-pressure chamber was recorded by a control device.

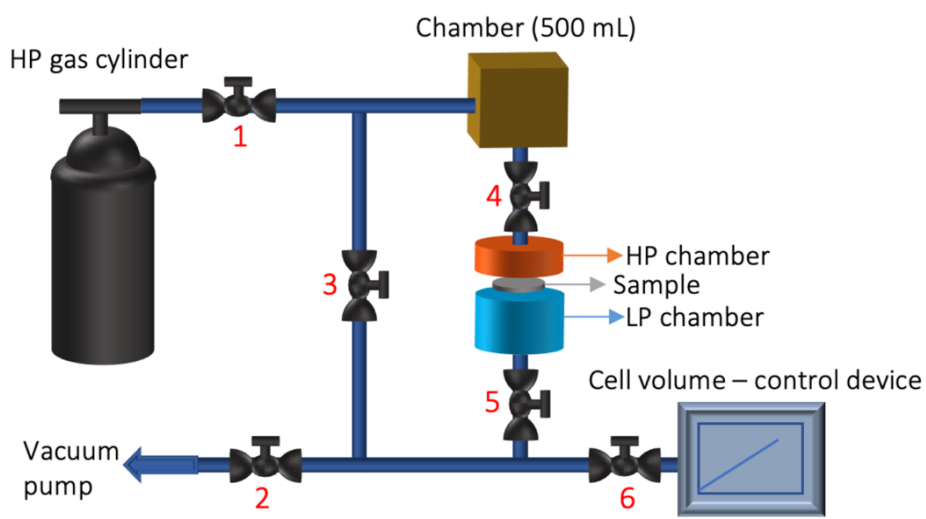

**Figure S3**. Schematic diagram of the $CO_2$ gas permeability rig, numbers 1-6 refer to the valves; HP and LP stand for high pressure and low pressure, respectively.

## S4. Statistics of the lateral size of the flakes

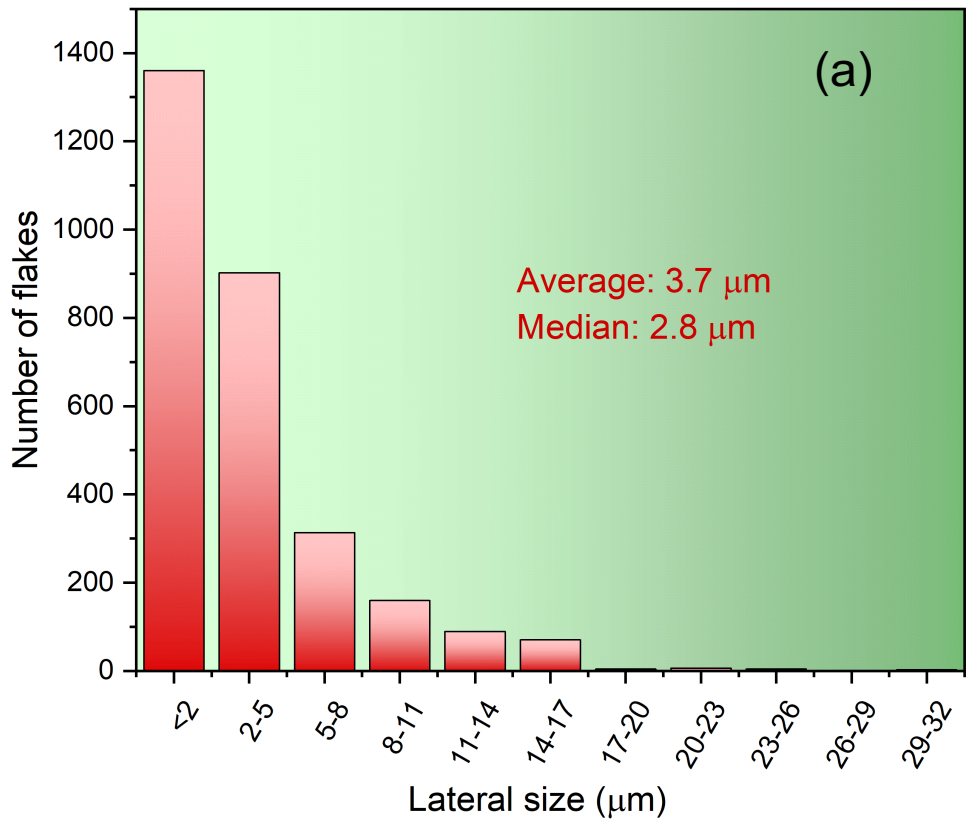

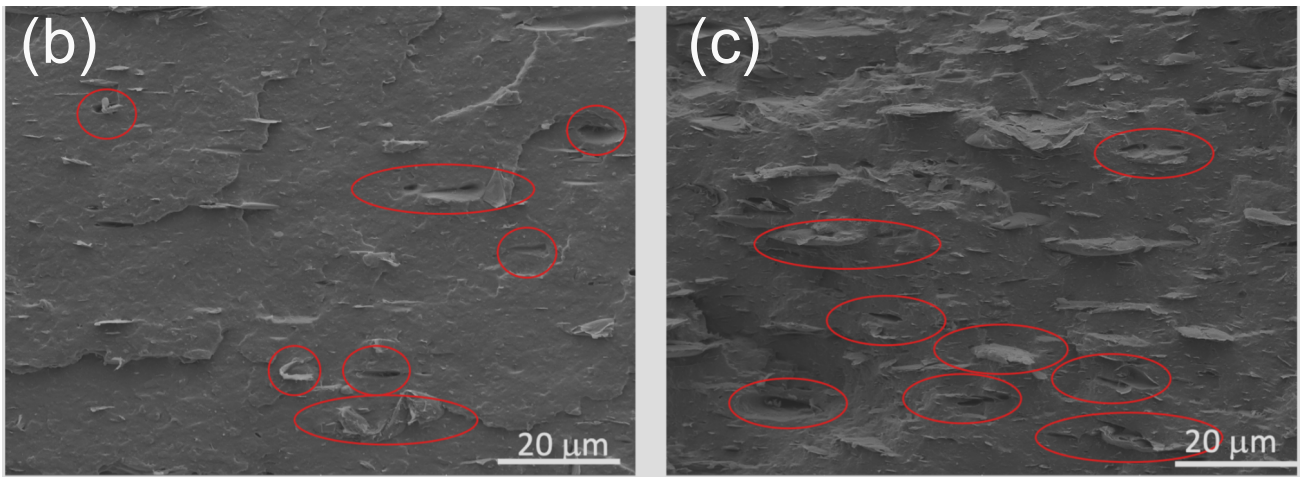

**Figure S4**. (a) Distribution of the lateral size of the flakes in the prepared composites; (b,c) loops and folds of flakes highlighted for samples of (b) GNP5 and (c) GNP10

## S5. $CO_2$ permeability

*Table S2.* The $CO_2$ permeabilities for all the samples tested.

| | FKM-GNP | | FKM-CB |
|---|---|---|---|
| phr | $P$ (×10$^{-15}$) (mol·m/(m$^2$·s·Pa)) | phr | $P$ (×10$^{-15}$) (mol·m/(m$^2$·s·Pa)) |
| 0 | 1.89 ± 0.19 | 0 | 1.89 ± 0.19 |
| 1 | 1.80 ± 0.25 | 15 | 1.40 ± 0.45 |
| 2.5 | 1.69 ± 0.29 | 30 | 1.17 ± 0.35 |
| 5 | 1.72 ± 0.21 | | |
| 10 | 1.35 ± 0.16 | | |
| 15 | 0.87 ± 0.17 | | |

## S6. Permeability modelling

In the Nielsen model, the permeability of a polymer filled with plate-like filler is given by [1],

$$P_c = P_m \left( \frac{1-V_f}{\tau} \right) \tag{S1}$$

where $P_m$ is the permeability of the neat polymer, the tortuosity factor is given by $\tau = \frac{d_a}{d_s} = 1 + \frac{s}{2} V_f$, $s$ is the aspect ratio of the platelets and $V_f$ is the volume fraction of the filler. In the paper of Nielsen [1], the derivation was not given. Hereby, the equation is derived, in order to clarify the prerequisite of using the equation for analysis. As can be seen in **Figure S4** (a), Nielsen's model defines a case of the tortuous path and thus the tortuosity factor $\tau$, that gives the maximized paths of a penetrate (gas molecule) throughout the material and therefore, the minimum permeability.

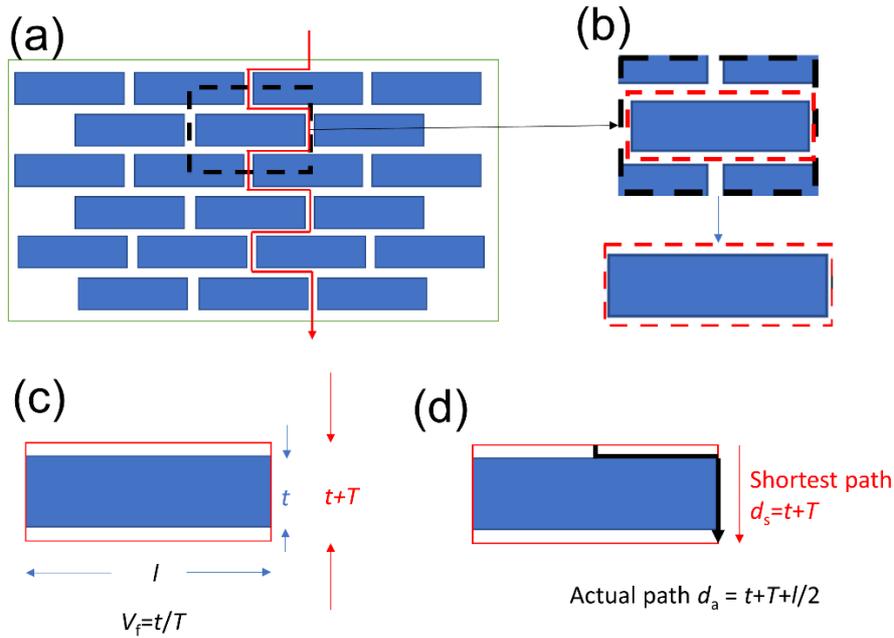

**Figure S5.** (a) Homogeneous platelets dispersed in a matrix and tortuous paths of a penetrate based on Nielsen's model with a selected representative volume element (RVE) (dashed box) for the composite; the red path with an arrow indicate the actual path of the penetrate. (b) Simplification of the RVE. (c) The RVE used in Nielsen's model, while $t$ is the thickness of the platelet, $T$ is the thickness of the surrounding matrix and $(t+T)$ is the thickness of the RVE; it is assumed that the end-to-end distance between flakes in the in-plane (horizontal) direction is infinite low; (d) the actual path ($d_a$) (black arrow) and the shortest path ($d_s$) (red arrow) of Nielsen's model.

The representative Nielsen's path (red path with an arrow) is shown in **Figure S5** (a), with a representative volume element (RVE) created in the black dashed box. From **Figure S5** (a) to (c), it shows the simplification of the RVE and finally the RVE in **Figure S5** (c) is the Nielsen's case. Based on the assumption that the end-to-end distance between flakes in the in-plane (horizontal) direction is infinite low, the volume fraction of the filler ($V_f$) is equal to $t/(t+T)$, where $t$ is the thickness of the flakes, and $(t+T)$ is the thickness of the RVE. Finally, in **Figure S5**(d), the actual distance ($d_a$) and the shortest distance ($d_s$) of the path is given by,

$$d_a = t + T + \frac{l}{2} \tag{S3}$$

and

$$d_s = t + T \tag{S4}$$

Therefore, the tortuosity factor $\tau$ is given by,

$$\tau = \frac{d_a}{d_s} = \frac{t + T + \frac{l}{2}}{t + T} = 1 + \frac{l}{2t} V_f \tag{S5}$$

where $l$ is the lateral size of the platelet and $t$ is the thickness of the platelets. Given that the aspect ratio $s$ is equal to $l/t$, combining equations (S1) and (S5), the permeability of a platelet-filled composite ($P_c$) is,

$$P_c = P_m \left( \frac{1 - V_f}{1 + \frac{s}{2} V_f} \right) \tag{S6}$$

Equation (S6) is the Nielsen's equation for permeability. It is important to note that the derivation is based on the assumption that the in-plane end-to-end distance between flakes is infinitesimally small. For the materials analysed in this work, where the filler loading is relatively low, the dispersion of the flakes can be schematically expressed as the case displayed in **Figure S6 (a)**, based on the SEM micrographs (**Figure 1**). It is apparent that this case in **Figure S6 (a)** fails to meet the infinite low end-to-end flake distance in the horizontal direction and therefore the transport path is unlikely to be maximized. For this reason, Nielsen's model is ineligible to be employed for the analysis of the permeability in this work. **Figure S6 (b)** presents the Nielsen's case with the same filler loading, where the dispersion of the flakes was subjectively set to meet the Nielsen's assumption. However, the dispersion shown in **Figure S6 (b)** is an ideal case, and it is very unlikely to be seen in actual samples with low filler loadings.

The dispersion we saw from the morphological characterisations can be analogous to **Figure S6** (a) rather than (b), and therefore the probability of the encounter between the penetrate and the flakes must be taken into account. In the following section, the derivation of the probability of a penetrate to encounter flakes is given.

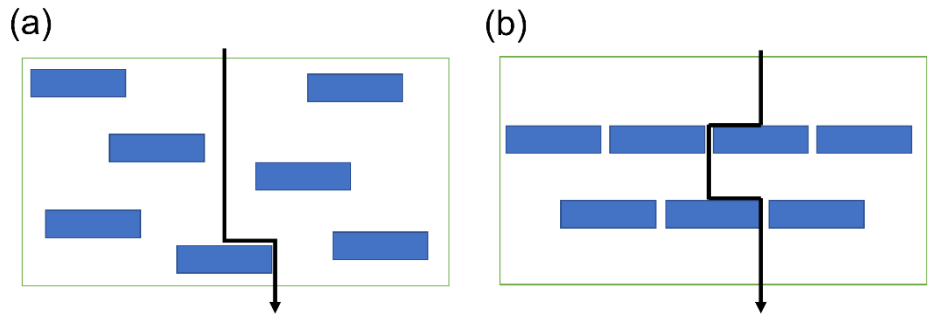

**Figure S6.** Dispersion of low loadings of flakes in the matrix: (a) homogeneous dispersion; (b) Nielsen's case.

The Binomial distribution is a mathematical approach that enables the analysis of the gas permeability of a composite filled by plate-like filler. The probability of a penetrate (gas molecule) to encounter $n$ flakes throughout $N$ layers can be obtained by a Binomial distribution [2],

$$P_n = \binom{N}{n} V_f^n (1-V_f)^{N-n} \tag{S7}$$

where

$$\binom{N}{n} \equiv \frac{N!}{n!(N-n)!} \tag{S8}$$

The function (S7) represents mathematically $n$ successes with $N$ trials, where each given trial has two possible outcomes: a 'success' with a probability of $V_f$ and a 'failure' with a probability of $(1-V_f)$. Each trial is an independent event. The permeability model studied in this paper is shown schematically in **Figure S7**. A trial is defined to be an event that a gas molecule enters and exit a layer (with a total number of layers throughout the composite defined as $N$). For each layer that a gas molecule penetrates, there is a probability of $V_f$ of an encounter between the gas molecule and a flake, and therefore there is a probability of ($1-V_f$) that the gas molecule misses a flake throughout the layer.

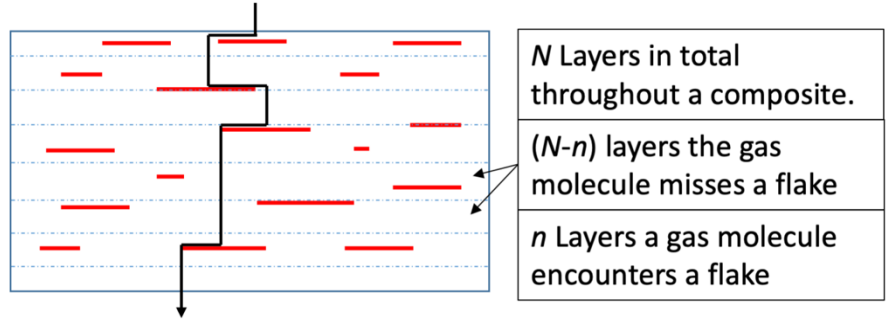

**Figure S7**. Schematic diagram of the path of gas transportation, where a size distribution of the flakes at low loadings of the filler is considered; the model enables the analysis for the gas molecules to miss a flake. This figure is reproduced from **Figure 5**(d).

For a random variable ($n$) with the Binomial probability distribution, the expected value (mean value) of $n$, that is $\bar{n}$, is given by,

$$\bar{n}=\sum_{n=0}^{N} n \frac{N!}{n!(N-n)!} V_f^n (1-V_f)^{N-n} \qquad (S9)$$

Since the term $n=0$ is equal to 0, equation S9 can therefore be rewritten into,

$$\bar{n}=\sum_{n=1}^{N} \frac{N!}{(n-1)!(N-n)!} V_f^n (1-V_f)^{N-n} \qquad (S10)$$

Let $x=n-1$ and $y=N-1$, then equation S10 is rewritten into,

$$\bar{n}=\sum_{x=0}^{y} \frac{(y+1)!}{x!(y-x)!} V_f^{x+1} (1-V_f)^{y-x} \qquad (S11)$$

and therefore,

$$\bar{n}=(y+1)V_f \sum_{x=0}^{y} \frac{y!}{x!(y-x)!} V_f^x (1-V_f)^{y-x} = NV_f \sum_{x=0}^{y} \frac{y!}{x!(y-x)!} V_f^x (1-V_f)^{y-x} \qquad (S12)$$

The Binomial theorem defines that,

$$(a+b)^y = \sum_{x=0}^{y} \frac{y!}{x!(y-x)!} \cdot a^x \cdot b^{y-x} \tag{S13}$$

Setting $a=V_f$ and $b=1-V_f$, then,

$$(a+b)^y = \sum_{x=0}^{y} \frac{y!}{x!(y-x)!} \cdot a^x \cdot b^{y-x} = \sum_{x=0}^{y} \frac{y!}{x!(y-x)!} V_f^x (1-V_f)^{y-x} = 1 \tag{S14}$$

Combining equations (S12) and (S14), we have,

$$\bar{n} = NV_f \text{ or } V_f = \frac{\bar{n}}{N} \tag{S15}$$